\documentclass[lettersize,journal]{IEEEtran}
\usepackage{cite}
\usepackage{amsmath,amssymb,amsfonts}
\usepackage{graphicx}
\usepackage{textcomp}

\usepackage{xcolor}
\usepackage{balance}
\usepackage{booktabs}
\usepackage{mathtools}
\usepackage{pifont}
\usepackage{epsfig,makeidx,color,subfigure}
\usepackage{amsmath,amssymb,bbm,enumitem}
\usepackage{cite,graphicx,lipsum}
\usepackage[switch,pagewise]{lineno}
\usepackage{tabularx}
\usepackage{hyperref}
\usepackage{algorithm,algpseudocode}

\hypersetup{
        colorlinks = false,
        citecolor = red,
        breaklinks = true,
        linktocpage=true,
}

\newcommand{\highlight}[2]{\colorbox{#1!17}{$\displaystyle #2$}}

\renewcommand{\highlight}[2]{\colorbox{#1!17}{#2}}

\def\bmu{\mbox{\boldmath $\mu$}}
\def\beps{\mbox{\boldmath $\epsilon$}}

\def\be{ \begin{equation} }
\def\ee{ \end{equation} }
\def\bea{ \begin{eqnarray} }
\def\eea{ \end{eqnarray} }

\def\bx{{\bf x}}
\def\by{{\bf y}}

\def\br{{\bf r}}

\def\bA{{\bf A}}
\def\bB{{\bf B}}

\def\bF{{\bf F}}

\def\bH{{\bf H}}

\def\b0{{\bf 0}}

\newtheorem{lemma}{Lemma}

\begin{document}

\title{Training-Free  Multi-User Generative Semantic Communications via Null-Space\\Diffusion Sampling}
\author{Eleonora~Grassucci,
Jinho~Choi~\IEEEmembership{Fellow,~IEEE}, Jihong~Park~\IEEEmembership{Senior Member,~IEEE}, Riccardo~F.~Gramaccioni~\IEEEmembership{Student Member,~IEEE}, Giordano~Cicchetti~\IEEEmembership{Student Member,~IEEE}, Danilo~Comminiello,
\IEEEmembership{Senior Member, IEEE}

\thanks{E. Grassucci, R. F. Gramaccioni, G. Cicchetti, and D. Comminiello are with the Dept. of Information Engineering, Electronics, and Telecommunications of Sapienza University of Rome, Italy. Emails: \{eleonora.grassucci, riccardofosco.gramaccioni, giordano.cicchetti, danilo.comminiello\}@uniroma1.it. J. Choi is with the School of Electrical and Mechanical Engineering, The University of Adelaide, Adelaide, Australia (e-mail: {jinho.choi@adelaide.edu.au}). J.~Park is with ISTD Pillar, Singapore University of Technology and Design, Singapore (e-mail: {jihong\_park@sutd.edu.sg}).
\\This work was partially supported by the European Union under the Italian National Recovery and Resilience Plan (PNRR) of NextGenerationEU, partnership on ``Telecommunications of the Future” (PE00000001 - program RESTART).}}

\maketitle

\begin{abstract}
Recent advances in artificial intelligence (AI) models, such as large language models and diffusion models, have shown significant potential in semantic communication by reconstructing multimedia data from highly compressed semantic signals under limited bandwidth and poor channel conditions. 
Unlike most existing approaches that focus on single-user scenarios with typical encoder-decoder models, this paper rethinks multi-user semantic communications using large generative models. In particular, in multi-user orthogonal frequency division multiple access (OFDMA) systems, we propose to reduce the number of subcarriers assigned per user by leveraging generative diffusion models to locally reconstruct missing or noisy information. By utilizing the null-space decomposition method for diffusion model sampling, we provide a traning-free, closed-form receiver design guideline for diffusion noise scheduling. Simulation results demonstrate that our proposed method achieves high-fidelity image reconstruction using only 60\% of the original subcarriers, and outperforms deep joint source-channel coding (DeepJSCC) and LDPC baselines by up to -10 dB in signal-to-noise ratio (SNR).
\end{abstract}


\section{Introduction}
\label{sec:introduction}

With the unprecedented growth in connected devices, ranging from mobile phones and cars to extended reality (XR) devices, there is an increasing demand for higher data rates \cite{zhijin2024ProcIEEE}. This necessitates new methods to support highly efficient multi-user communication within limited bandwdith. 
Meanwhile, a novel paradigm has emerged laying its foundations in the three Weaver levels of communication. Here, the first level is the \textit{technical level}, whose aim is to manage the technical aspects of the transmission. The \textit{semantic level} is placed right after the first one and focuses on understanding \textit{what} to transmit rather than \textit{how} to do it. Finally, the last one regards the \textit{effectiveness} of the communication. Therefore, the so-called semantic communication was born from the middle level and aims at transmitting and reconstructing the meaning (i.e., the \textit{semantics}) of the message without necessarily recovering the original bitstream \cite{Mahgoub2026Access, zhijin2024ProcIEEE, Nan2023JSAC, Dai2021CommunicationBT, Choi2024SemanticCC}. Therefore, semantic communication frameworks can reduce bandwidth requirements and latency by transmitting only the key semantic information of the content.

\begin{figure}
    \centering
    \includegraphics[width=\linewidth]{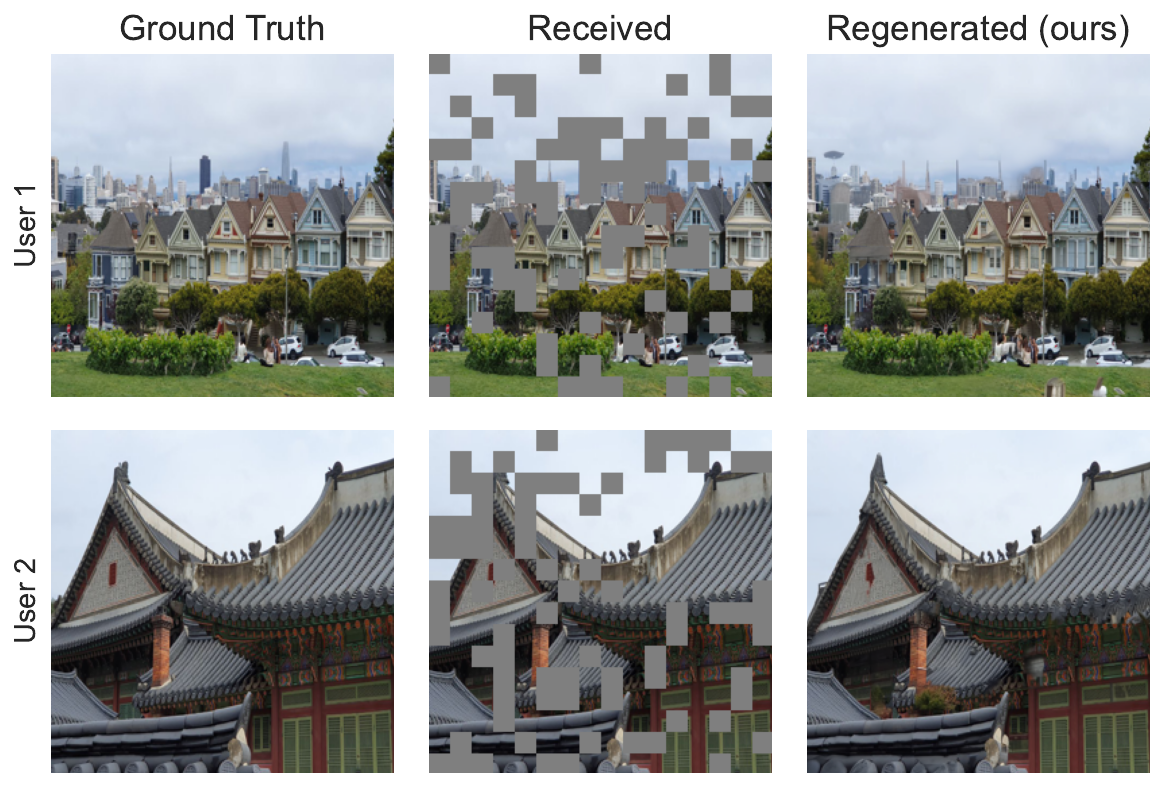}
    \caption{Sample results of our method on out-of-dataset images for two-user scenario and number of subcarriers $N < M$, with $M = 256$ and $N = 180$, equivalent to an $N/M$ ratio equal to $0.7$. The proposed method effectively inpaints the gaps in the received information.}
    \label{fig:fig1}
\end{figure}

Concurrently, semantic communication has been discovered to be a fertile layout for involving large generative models in communication frameworks \cite{Grassucci2023EnhancingSC, Wu2025TWC, grassucci2024generative, Guo2024TWC}. Such models are among the most impressive and promising branches of artificial intelligence, possessing the ability to generate almost any kind of multimedia content ranging from text (large language models) to audio and video (diffusion models). Their key feature for semantic communication is the ability to generate content including images \cite{Sun2025Access, Xu2025JSAC, Grassucci2023GenerativeSC, Zhang2025TCCN} and audio \cite{Grassucci2023DiffusionMF} or speech \cite{Han2022SemanticPreservedCS, Jang2024ICASSP} from extremely compressed information such as text \cite{Nam2023LanguageOrientedCW, Nam2023SECON} or lower-dimensional and quantized data representations \cite{Nemati2023}, further reducing the amount of information to transmit in a communication system and expanding the range of possible applications \cite{grassucci2024generative}. The combination of the revolutionary ability of generative models with the semantic communication paradigm gave birth to the generative semantic communication research field \cite{Xu2025JSAC, Grassucci2023GenerativeSC, Nam2023LanguageOrientedCW, Pei2025TWC}, which is showing promising and revolutionary results. However, none of these methods consider the multi-user scenario yet, in which conventional deep learning models have instead shown interesting performance \cite{Xie2022JSAC, Loc2024TVT, Xu2023TCOMM}.
Most of the generative semantic communication approaches focus on inserting generative models in the semantic level of conventional communication frameworks, leveraging their capabilities to reduce bandwidth requirements and latency while improving perceptual results at the receiver.

In this paper, we propose to rethink multi-user semantic communications empowering the system with large generative models and assigning resources and communication channels to multiple users given the knowledge that the system can exploit generative models ability to regenerate missing portions of the message at each receiver. Therefore, under the proposed new paradigm, in the case of multi-user congestion, the channel-user assignment should not aim at receiving the largest part of information implying multiple transmissions over time. Rather, it should be limited to transmitting the sufficient information required to the generative model for regenerating the missing content. More technically, we propose to express the problem of multi-user semantic communication as an inverse problem and leverage the capabilities of diffusion models to solve it \cite{Moliner2022SolvingAI, murata2023gibbsddrm, Meng2021SDEditGI}.
To solve the inverse problem, we formulate the diffusion model sampling algorithm according to the null-space decomposition theorem \cite{Schwab_2019} to precisely and formally match the generative algorithm with the scenario of multi-user semantic communication. The proposed sampling strategy can be adopted with any pretrained diffusion model, making it extremely flexible to be involved in any pre-existing generative semantic communication framework based on diffusion models.
Following the proposed method, generative models can play a key role in reshaping semantic communication systems in multi-user scenarios towards a GenAI-based next generation of communications. 

Our contributions are four-fold as follows:
\begin{itemize}
    \item We propose a novel method to rethink multi-user semantic communications leveraging the potential of large generative models. The proposed sampling method can be plugged into any existing pretrained generative model.
    \item We provide a formulation that matches the muli-user semantic communication problem and the proposed diffusion model generative algorithm.
    \item We solve the multi-user Orthogonal Frequency Division Multiple Access (OFDMA) problem avoiding retransmission or unreceived content by designing an effective diffusion model equipped with a novel null-space decomposition sampling method.
    \item We prove the effectiveness of the proposed approach in multiple scenarios, across different numbers of subcarriers and channel noises.
\end{itemize}

The rest of the paper is organized as follows. Section~\ref{sec:related_works} reports the works related to semantic communication and diffusion models, while Section~\ref{sec:background} sets the problem and explains the theory behind diffusion models. Then, the proposed method is introduced in Section~\ref{sec:method} and validated in Section~\ref{sec:exp}, while conclusions are drawn in Section~\ref{sec:con}.

\section{Related Works}
\label{sec:related_works}


\subsection{Semantic Communication}

Semantic communication is expected to play a key role in future communication systems beyond 5G and 6G \cite{Luo2022WC, Huang2023JointTA, Qin2021SemanticCP, Sun2025Access}. 
The idea behind this new paradigm is to focus on transmitting the semantics of message or data, which is expected to contain the meaning and key features of the original content, rather than the entire original content. Such semantic information is often very small in size and invariant to perturbation as compared with its original data. Then, the receiver can leverage the semantic information to restore or regenerate the message, or directly involve the semantics to accomplish specific goals or perform certain tasks \cite{strinati2024goaloriented}. Under this presupposition, semantic communication frameworks are expected to reduce the bandwidth required for the transmission and improve robustness in poor channels. As a result, semantic communication systems could be applied to diverse applications, ranging from image transmission \cite{Patwa2020SemanticPreservingIC, Wang2019AnED}, to speech \cite{Weng2021ICC, Xiao2023ICASSP, Han2022SemanticPreservedCS}, video compression and transmission \cite{Jiang2022WirelessSC, Guo2025JSAC}, and it is expected to explode in much more fields of applications \cite{Dai2021CommunicationBT, grassucci2024generative}.
Concurrently, interest has been growing around token semantic communication, which leverages the power of large transformer models to transmit and recover semantic patches of images \cite{devoto2024globecom} and that has been applied to multi-user semantic communication \cite{qiao2025tokendomainmultipleaccessexploiting}. However, unlike the method in \cite{qiao2025tokendomainmultipleaccessexploiting}, where the restoration relies entirely on the pre-trained foundation model capability, our proposed method showcases that it is promising to utilize not only the pretrained foundation model capability (diffusion model) but also additional information on the perturbation and channel (i.e., masking or subcarrier-allocation matrix A and noise variance). 

\subsection{Diffusion Models}

Denoising diffusion probabilistic models \cite{ho2020denoising} (diffusion models, in short) have recently become the state of the art for generating multimedia content ranging from images \cite{saharia2022photorealistic, Rombach2021latent} and audio \cite{Liu2023AudioLDM2L, Ghosal2023TexttoAudioGU} to video \cite{huang2025planxinstructvideogeneration, Jiang2023Text2PerformerTH}, usually conditioned on some user-friendly representation such as textual description \cite{saharia2022photorealistic, huang2025planxinstructvideogeneration}. We can identify two key elements responsible for diffusion models success and widespread. First, the generation ability of diffusion models crucially outperforms the capabilities of other models, such as Generative Adversarial Networks (GANs) or Variational Autoencoders (VAEs) \cite{Dhariwal2021BeatGANs}. 
Second, the sampling process of diffusion models is far more stable than the generation process of GANs \cite{Croitoru2023TPAMI}, making it a more reliable method compared to them.
Together, these aspects also favor the adoption of diffusion models in semantic communication scenarios \cite{Liu2025TCCN, Xu2025JSAC, Grassucci2023GenerativeSC, Grassucci2023DiffusionMF, Yang2024ICASSP, Pei2025TWC} or multi-user ones \cite{zeng2024dmce}, whose models reliability is crucial and where their ability to regenerate content from extremely compressed information perfectly fits the semantic communication scenario. Indeed, generative semantic communication frameworks based on diffusion models have clearly outperformed existing methods based on GANs \cite{Gunduz2022GenSem} and VAEs \cite{Malur2020VAE, Estiri2020AVA, Liang2026Survey}.


\section{Problem Formulation}
\label{sec:background}

In this Section, we present in detail the problem formulation for applying semantic communication to multiuser systems.


Consider a system comprising a base station and $K$ users for downlink transmission. Let $\mathbf{x}_k \in \mathbb{C}^M$ denote the signal (represented as a vector of length $M$) to be received by user $k$. The transmitted signal by the base station through the downlink channel is given by

\vspace{6pt}

\begin{equation}
    \mathbf{y} = \sum_{k=1}^K \mathbf{F}_k \mathbf{x}_k \in \mathbb{R}^L,
\label{EQ:yBx}
\end{equation}
\normalsize
where $\mathbf{F}_k \in \mathbb{C}^{L\times M}$ is the channel assignment matrix to user $k$ and $L \ge M$. Here, a radio resource block of length $L$ is used to transmit $K$ users' signals of length $M$. Throughout the paper, we assume OFDMA \cite{GoldsmithBook}. Thus, each element of $\by$ is transmitted through a subcarrier. Note that it is necessary to have 
\be 
\bF_k^\mathrm{H} \bF_{l} = \b0, \quad k \ne l,
    \label{EQ:BB}
\ee 
to ensure orthogonal channel allocations.
 
Let $\mathbf{H}_k$ be the $L \times L$ channel matrix of user $k$, which is diagonal in OFDMA. That is, the $l$th diagonal element of $\bH_k$ is the (frequency-domain) channel coefficient of subcarrier $l$ from the base station to user $k$. 
It is also assumed that Time Division Duplex (TDD) is used so that the channel matrices are known at the transmitter (i.e., the base station), thanks to the channel reciprocity. Then, the received signal at user $k$ over channel $\mathbf{H}_k$  is
given by
\begin{align}
    \mathbf{r}_k 
    & = \mathbf{H}_k \mathbf{y} + \mathbf{n}_k \cr 
    & = \mathbf{H}_k \mathbf{F}_k \mathbf{x}_k + \mathbf{H}_k \sum_{l \neq k} \mathbf{F}_l \mathbf{x}_l + \mathbf{n}_k,
\label{EQ:rk}
\end{align}
where $\mathbf{n}_k$ is the background noise at user $k$, for simplicity we consider additive white Gaussian noise. 

If the base station wishes to send all the signals with orthogonal channel allocations, as shown in eq.~\eqref{EQ:BB}, it can be shown that 
$$
L = K M.
$$
That is, in OFDMA, in order to support $K$ users with signal vectors to be transmitted over $M$ subcarriers, there should be a total of $L = K M$ subcarriers.  To potentially reduce the number of necessary subcarriers, two key factors should be considered:
\begin{itemize}
\item[{\bf F1)}] Due to varying multipath fading among users, the channel matrices differ, and certain subcarriers of specific users may experience deep fading. Therefore, as assumed previously, since the base station possesses knowledge of the channel matrices, it can allocate channels to circumvent deep fading.
\item[{\bf F2)}] In addition to channel-adaptive allocations, it is possible to reduce the dimension of the signal using large generative models.
\end{itemize}

Taking into account the aforementioned factors, we assume that each user is allocated $N$ subcarriers, where $N < M$. Let $\mathbf{B}_k \in \mathbb{C}^{L \times M}$ represent the reduced-dimension channel allocation matrix of user $k$. This matrix can be considered a submatrix of $\mathbf{F}_k$. 
Moreover, it is crucial to note that these channel matrices are determined based on the channel conditions of each user, aiming to avoid deep fading. In essence, each user can only utilize subcarriers with sufficiently high channel gains. Furthermore, to ensure orthogonal allocations, the following condition must hold:
\be 
\bB_k^\mathrm{H} \bB_{l} = \b0, \quad k \ne l.
    \label{EQ:BB2}
\ee 
Therefore, a total of $L = KN$ (not $KM$) subcarriers are required for downlink transmissions. 

Since $N < M$, it is clear that some parts of the signal, $\bx_k$, cannot be transmitted. These missing parts can be generated using large generative models. To this end, let $\mathbf{A}_k = \mathbf{H}_k \mathbf{B}_k$. Then, we have
\begin{equation}
    \mathbf{x}_k = \mathbf{A}_k^\dagger \mathbf{A}_k \mathbf{x}_k + (\mathbf{I} - \mathbf{A}_k^\dagger \mathbf{A}_k) \mathbf{x}_k .
\label{EQ:xx}
\end{equation}
At user $k$, $\bA_k$ is known (as the reduced-dimension channel allocation matrix, $\bB_k$, and the channel matrix, $\bH_k$, are known), and from eq.~\eqref{EQ:rk}, $\bA_k \bx_k$ (with the background noise) can be obtained. Due to the orthogonal channel allocations in eq.~\eqref{EQ:BB2}, the signals to the other users can be removed by applying $\mathbf{A}_k^\dagger$ to $\br_k$. In other words, 
\be 
\mathbf{A}_k^\dagger \mathbf{r}_k =  \mathbf{A}_k^\dagger \mathbf{A}_k \mathbf{x}_k + \mathbf{A}_k^\dagger \mathbf{n}_k.
\ee 
If a generative model at user $k$ is able to generate the \highlight{orange}{missing parts of the signal}, i.e., $(\mathbf{I} - \mathbf{A}_k^\dagger \mathbf{A}_k) \mathbf{x}_k $, then from eq.~\eqref{EQ:xx}, an estimate of $\mathbf{x}_k$ can be given by
\begin{align}
\label{eq:problem}
    \hat{\mathbf{x}}_k 
    & = \highlight{teal}{$\mathbf{A}_k^\dagger \mathbf{r}_k$} + \highlight{orange}{$(\mathbf{I} - \mathbf{A}_k^\dagger \mathbf{A}_k) \tilde{\mathbf{x}}_k$} \cr 
    & = \mathbf{A}_k^\dagger \mathbf{A}_k \mathbf{x}_k + (\mathbf{I} - \mathbf{A}_k^\dagger \mathbf{A}_k) \tilde{\mathbf{x}}_k + \mathbf{A}_k^\dagger \mathbf{n}_k  \cr
    & = \bx_k + (\mathbf{I} - \mathbf{A}_k^\dagger \mathbf{A}_k)  (\tilde{\mathbf{x}}_k - \mathbf{x}_k) +\mathbf{A}_k^\dagger \mathbf{n}_k ,
\end{align}
where $\tilde{\mathbf{x}}_k$ is a generated signal at the receiver, i.e., user $k$, which is resposible to provide
the missing part of the signal, $(\mathbf{I} - \mathbf{A}_k^\dagger \mathbf{A}_k) \mathbf{x}_k$.

Of course, as the ratio $\frac{N}{M}$ decreases, it is expected that the estimation error may increase, despite the potential for achieving higher spectral efficiency. Yet, it may still be possible to recover the signal reasonably well in terms of semantic-related metrics, even with a significant loss.

\begin{figure}
    \centering
    \includegraphics[width=\linewidth]{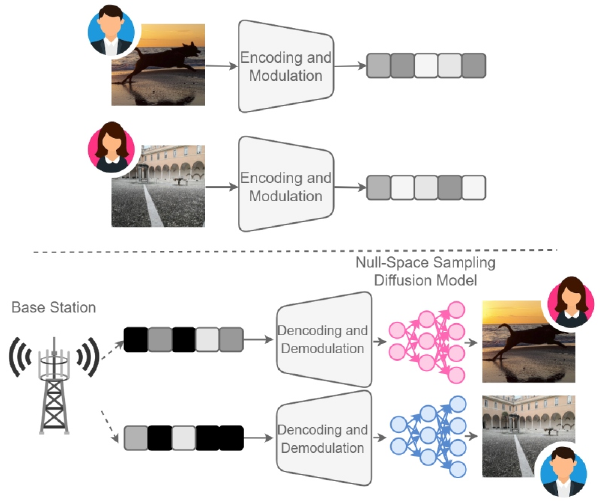}
    \caption{Schematic representation of the proposed generative semantic communication system with the proposed null-space diffusion model to regenerate missing portions of the bitstream (black dots in the figure) at the receivers.}
    \label{fig:system}
\end{figure}


\section{Solving multi-user semantic communication with diffusion models}
\label{sec:method}

In this section, we propose the generative semantic communication system to support multiple users, building upon the problem formulation introduced in Section~\ref{sec:background}. A simplified example of the system is also shown in Fig.~\ref{fig:system}.
We apply a diffusion model to recover missing portions of the signal. Specifically, we utilize an OFDMA system for downlink transmissions, where each user is allocated a smaller number of subcarriers ($N$) to transmit a signal of length $M$, where $M > N$. 
Consequently, certain parts of each user's signal are missing, which the proposed null-space diffusion sampling at the receivers aims to fill in. In Sec. \ref{subsec:multiuser}, we first describe how the $K$-user communication problem is recast as a single-user null-space diffusion sampling process. This is viable as interference from other users can be mitigated, as demonstrated in eq.~\eqref{eq:problem}, thanks to orthogonal channel allocations. Then, in Sec.~\ref{subsec:method}, we will omit the user index $k$.


\subsection{Denoising Diffusion Probabilistic Model}    \label{SS:Diff}


The core structure of diffusion models \cite{ho2020denoising} is a Markov chain that goes from time $0$ to time $T$ in the forward process, denoted by $q$, and from the time step $T$ to $0$ in the reverse process, denoted by $p$. On one hand, the forward direction starts from an image and progressively adds white Gaussian noise to destroy all the information in the image at time $T$. On the other hand, the reverse process slowly builds the desired data $\mathbf{x}_0$ from the original noise sample $\mathbf{x}_T$. The transition probability in the forward process is normally-distributed as

\small
\begin{equation}
    q(\mathbf{x}_t | \mathbf{x}_{t-1}) = \mathcal{N}(\sqrt{1-\beta_t} \mathbf{x}_{t-1}+\sqrt{\beta_t}\beps; \sqrt{1-\beta_t} \mathbf{x}_{t-1}, \beta_t \mathbf{I}),
\end{equation}
\normalsize
\noindent where $\epsilon \sim \mathcal{N}(\mathbf{0}, \mathbf{I})$, $\mathbf{x}_t$ is the intermediate noisy image at time $t$, and $\beta_t$ is the pre-defined variance schedule. By reparameterizing with $\alpha_t = 1-\beta_t$ and $\bar{\alpha}_t = \prod_{i=0}^t \alpha_i$, the forward process formulation becomes:
\begin{equation}
\label{eq:forward}
    q(\mathbf{x}_t | \mathbf{x}_0) = \mathcal{N}(\mathbf{x}_t; \sqrt{\bar{\alpha}_t}\mathbf{x}_0, (1-\bar{\alpha}_t) \mathbf{I}).
\end{equation}

Instead, the posterior distribution of the transition probabilities of the reverse process is derived from the forward process equations by applying the Bayes theorem as:

\begin{align}
    p(\mathbf{x}_{t-1}|\mathbf{x}, \mathbf{x}_0) 
    &= q(\mathbf{x}_t|\mathbf{x}_{t-1}) \frac{q(\mathbf{x}_{t-1}|\mathbf{x}_0)}{q(\mathbf{x}_t|\mathbf{x}_0) }\\ 
    &= \mathcal{N}(\mathbf{x}_{t-1}; \bmu_t(\mathbf{x}_t, \mathbf{x}_0), \sigma^2_t \mathbf{I}),
\end{align}
where the mean and the variance have the following forms:
\begin{align}
    \bmu_t(\mathbf{x}_t, \mathbf{x}_0) &= \frac{1}{\sqrt{\alpha_t}} \left(\mathbf{x}_t - \epsilon \frac{1-\alpha_t}{\sqrt{1-\bar{\alpha}_t}} \right) \\
    \sigma^2_t &= \frac{1-\bar{\alpha}_{t-1}}{1-\bar{\alpha}_t}\beta_t.
\end{align}

Under this construction, diffusion models involve a denoising neural network $\mathcal{Z}_\theta$ in the forward process that is trained to predict the noise at the given time step. More precisely, the noise $\beps \sim \mathcal{N}(\mathbf{0}, \mathbf{I})$ is applied at a randomly-sampled time step $t$ to the original image $\mathbf{x}_0$, which will be the input to the network $\mathcal{Z}_\theta$ to update its parameters $\theta$.
Once the training is finished, the denoising network can be exploited to progressively denoise the noise sample $\mathbf{x}_T$ up to generate the new image $\mathbf{x}_0$.

Therefore, the whole model is trained to let the network match the noise $\epsilon$ by minimizing the following loss function:
\begin{equation}
    \mathcal{L} = \| \epsilon - \mathcal{Z}_\theta(\sqrt{\bar{\alpha}_t}\mathbf{x}_0 + \epsilon \sqrt{1- \bar{\alpha}_t}, t) \|_2^2.
\end{equation}

\subsection{ Multi-User Communication via Inverse Generation} \label{subsec:multiuser}

In multi-user OFMDA communication with $K$ users and a total of $L=KN$ subcarriers, suppose that user $k$ downloads a length $M$ signal $\mathbf{x}_k$ using only $N<M$ subcarriers. Here, we consider that $N$ subcarriers are selected based on the $N$ highest gains in $\mathbf{H}_k$ while ensuring the orthogonal allocation constraint in eq.~\eqref{EQ:BB2}. In this case, the received signal $\mathbf{r}_k=\mathbf{H_k}\mathbf{B}_k \mathbf{x}_k+ \mathbf{n}_k $ at user $k$ results from the original signal $\mathbf{x}_k$ being perturbed by: linear scaling $\mathbf{H}_k$ and masking $\mathbf{B}_k$ (i.e., $\mathbf{A}_k$) as well as additive noise $\mathbf{n}_k$. Therefore, reconstructing $\mathbf{x}_k$ from $\mathbf{r}_k$ boils down to the problems of linear inverse scaling, inpainting, and denoising, respectively. These three problems can be combined as an inverse generation problem using a diffusion model. 

The standard diffusion models are compatible with real values, whereas $\mathbf{x}_k\in\mathbb{C}^M$ and its subsequent parameters are complex-valued, as defined in Sec. \ref{sec:background}. To address this mismatch, we hereafter consider that $\mathbf{x}_k = [\mathbf{x}_{k,R}^\text{T}, \mathbf{x}_{k,I}^\text{T}]^\text{T}  \in\mathbb{R}^M$ where $\mathbf{x}_{k,R}\in \mathbb{R}^\frac{M}{2}$ and $\mathbf{x}_{k,I}\in \mathbb{R}^\frac{M}{2}$ are the real and imaginary parts of a complex signal in $\mathbb{C}^\frac{M}{2}$. User $k$, equipped with a diffusion model, aims to reconstruct $\mathbf{\hat{x}}_k \approx \mathbf{x}_k$ from the received $\mathbf{r}_k$. To this end, the diffusion model at user $k$ generates $\mathbf{\hat{x}}_k$ through $T$ diffusion sampling steps. At each $t$-th sampling step, the denoising parameters are optimized for given $\mathbf{A}_k$ and $\mathbf{n}_k$, and the generated sample is combined with $\mathbf{r}_k$ while satisfying its desired range and null space characteristics. This process can be implemented by modifying the null-space diffusion sampling technique \cite{Wang2022ZeroShotIR, Grassucci2023DiffusionMF}. The aforementioned solution can be applied to any user $k$ under OFDMA owing to orthogonal channel allocations. Therefore, we henceforth focus only on a single user's null-space diffusion sampling process while omitting the user index $k$, as we shall elaborate in Sec.~\ref{subsec:method}.





\subsection{ Null-Space Diffusion Sampling for Inverse Generation}
\label{subsec:method}

\begin{figure*}
    \centering
    \includegraphics[width=\textwidth]{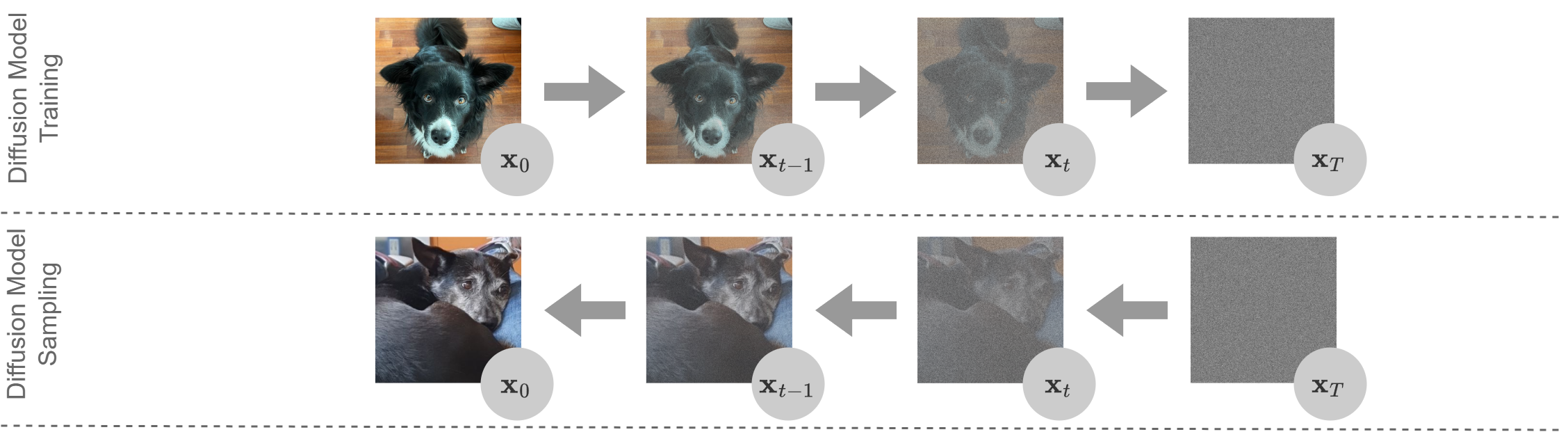}
    \includegraphics[width=\textwidth]{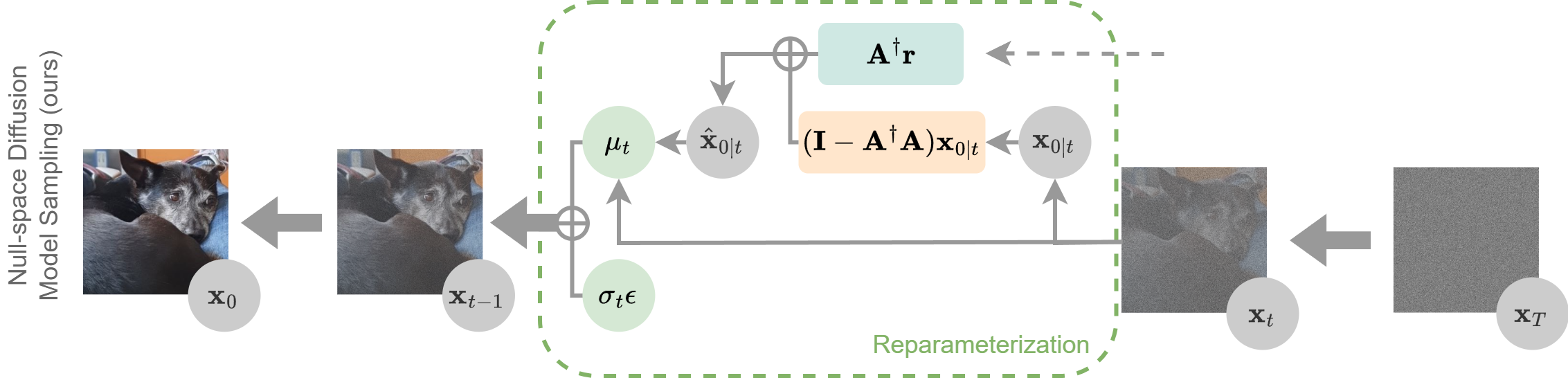}
    \caption{Illustration of diffusion model forward process, utilized in training, diffusion model standard reverse process for sampling, and ours null-space diffusion model sampling with the reparameterization of the mean and variance.}
    \label{fig:ourmethod}
\end{figure*}

From the null-space decomposition \cite{Schwab_2019}, we know that, for a matrix $\mathbf{A}^{n \times n}$:

\noindent \textbf{Definition 1} \textit{The range space of $\mathbf{A}$, $R(\mathbf{A})$, is the subspace spanned by the column of $\mathbf{A}$. The null space of $\mathbf{A}$, $N(\mathbf{A})$, is the solution space of the linear system $\mathbf{A}\mathbf{x}=0$}.

From the Dimension Theorem, we also know that $\text{dim} R(\mathbf{A}) + \text{dim} N(\mathbf{A}) = n$, and that if $\mathbf{A}$ is non-singular, then $N(\mathbf{A}) = 0$ and $R(\mathbf{A}) \cap N(\mathbf{A}) = 0$. If $\mathbf{A}$ is singular, then it may be possible that $N(\mathbf{A}) = 0$ and $R(\mathbf{A}) \cap N(\mathbf{A}) \neq 0$ but it can be proven that the null-space decomposition holds the same.

The operator $\mathbf{A}^\dagger \mathbf{A}$ projects $\mathbf{x}$ in the range space of $\mathbf{A}$ since $\mathbf{A} \mathbf{A}^\dagger \mathbf{A} \mathbf{x} = \mathbf{A} \mathbf{x} = \mathbf{r}$. Instead, the operator $(\mathbf{I}-\mathbf{A}^\dagger \mathbf{A})$ projects $\mathbf{x}$ in the null space of $\mathbf{A}$ due to $\mathbf{A}(\mathbf{I} - \mathbf{A}^\dagger \mathbf{A})\mathbf{x} = 0$.

\noindent Following this formulation, any sample $\mathbf{x}$ can be decomposed in the range space and the null space as:

\begin{equation}
    \mathbf{x} = \mathbf{A}^\dagger \mathbf{A} \mathbf{x} + (\mathbf{I}-\mathbf{A}^\dagger \mathbf{A}) \mathbf{x}.
\end{equation}

Considering a generic inverse problem $\mathbf{r} = \mathbf{A} \mathbf{x}$, the solution to this problem $\hat{\mathbf{x}}$ must satisfies two constraints:

\begin{align}
    \text{\textit{Consistency}:} \quad &\mathbf{A}\hat{\mathbf{x}}=\mathbf{r} \\
    \text{\textit{Realness}:} \quad &\hat{\mathbf{x}} \sim q(\mathbf{x}).
\end{align}

\noindent The solution $\hat{\mathbf{x}}$ that satisfies the consistency constraint is:

\begin{equation}
\label{eq:consistency_sat}
    \hat{\mathbf{x}} = \highlight{teal}{$\mathbf{A}^\dagger \mathbf{r}$} + \highlight{orange}{$(\mathbf{I}- \mathbf{A}^\dagger \mathbf{A}) \tilde{\mathbf{x}}$},
\end{equation}

\noindent which recalls the formulation of the multi-user semantic communication problem in eq.~\eqref{eq:problem}, except for the noisy element that we will introduce later in this Section. However, while the formulation in eq.~\eqref{eq:consistency_sat} satisfies the consistency constraint, the element that instead controls the realness constraint is $\tilde{\mathbf{x}}$. Therefore, the objective of the diffusion model training is \highlight{orange}{generating the proper null space $(\mathbf{I} - \mathbf{A}^\dagger \mathbf{A} )\tilde{\mathbf{x}}$} according to the \highlight{teal}{range space $\mathbf{A}^\dagger\mathbf{r}$} such that the realness constraint is satisfied, i.e., $\hat{\mathbf{x}}$ comes from the data distribution $q(\mathbf{x})$. Visually, the proposed sampling method is shown in Fig. \ref{fig:ourmethod}.
The following Lemma \ref{lem:range-null} shows the equivalence between the formulation in eq.~\ref{eq:consistency_sat} and the one in eq.~\eqref{eq:problem}.

\begin{lemma}[\textbf{OFDMA/Null-Space Formulation Equivalence}]\label{lem:range-null}
For any user~$k$ with channel allocation matrix
$\mathbf{A}_k \in \mathbb{C}^{N\times M}$:


\[
\hat{\mathbf{x}}_k \;=\;
\underbrace{\highlight{teal}{$\mathbf{A}_k^{\dagger}\mathbf{r}_k$}}_{\substack{\text{\textbf{range space}}\\[-1pt]\text{\textbf{(received by transmission)}}}} 
\;+\;
\underbrace{ \highlight{orange}{$(\mathbf{I}-\mathbf{A}_k^{\dagger}\mathbf{A}_k)\tilde{\mathbf{x}}_k$}}_{\substack{\text{\textbf{null space}}\\[-1pt] \text{\textbf{(missing, to be generated)}}}}.
\]

\end{lemma}

\textbf{Implication.} For each user $k$, the diffusion model can generate the null-space term, i.e. the missing parts at the receiver, while the range-space term is received through transmission.

However, additional considerations to eq.~\ref{eq:consistency_sat} should be evaluated when designing null-space diffusion models \cite{Wang2022ZeroShotIR, Grassucci2023DiffusionMF}.
Indeed, as stated in Section \ref{sec:background}, when sampling $\mathbf{x}_{t-1}$ from $p(\mathbf{x}_{t-1}|\mathbf{x}_t, \mathbf{x}_0)$ in the reverse process, where $\mathbf{x}_0$ {denotes the original image, the intermediate states $\mathbf{x}_t$  is perturbed by Gaussian noise, making the null-space decomposition unfeasible. Therefore, we have to introduce a reparameterization to obtain noise-free intermediate states and apply the null-space decomposition. Precisely, we reparameterize the mean $\bmu_t(\mathbf{x}_t, \mathbf{x}_0)$ and the variance $\sigma_t^2$ of the reverse process transition distribution $p(\mathbf{x}_{t-1}|\mathbf{x}_t, \mathbf{x}_0)$ as:

\begin{align}
\label{eq:meanvar}
    \bmu_t(\mathbf{x}_t, \mathbf{x}_0) &= \frac{\sqrt{\bar{\alpha}_{t-1}}\beta_t}{1-\bar{\alpha}_t}\mathbf{x}_0 + \frac{\sqrt{\alpha_t}(1-\bar{\alpha}_{t-1})}{1-\bar{\alpha}_t} \mathbf{x}_t \\
    \sigma^2_t &= \frac{1-\bar{\alpha}_{t-1}}{1-\bar{\alpha}_t}\beta_t.
\end{align}

Although the noise-free image $\mathbf{x}_0$ is still unknown in intermediate states, we can obtain an estimate $\mathbf{x}_{0|t}$ of it from $\mathbf{x}_t$ by reversing eq.~\eqref{eq:forward} and predict the noise with the denoising model $\epsilon_t = \mathcal{Z}_\theta(\mathbf{x}_t, t)$, i.e.:

\begin{equation}
    \mathbf{x}_{0|t} = \frac{1}{\sqrt{\bar{\alpha}_t}}(\mathbf{x}_t - \sqrt{1- \bar{\alpha}_t} \mathcal{Z}_\theta(\mathbf{x}_t, t)).
\end{equation}

\noindent To obtain the final estimate of the noise-free intermediate information, we can fix the range space, yielding:

\begin{equation}
\label{eq:x0t}
    \hat{\mathbf{x}}_{0|t} = \mathbf{A}^\dagger \mathbf{r} + (\mathbf{I} - \mathbf{A}^\dagger \mathbf{A}) \mathbf{x}_{0|t} = \mathbf{x}_{0|t} - \mathbf{A}^\dagger (\mathbf{A} \mathbf{x}_{0|t}- \mathbf{A}\mathbf{x}).
\end{equation}

\noindent We employ the fresh estimation of eq.~\eqref{eq:x0t} in the reparameterization of the mean and of the variance in eq.~\eqref{eq:meanvar}, thereby yielding the intermediate sample $\mathbf{x}_{t-1}$ as:

\begin{equation}
\label{eq:interstate}
    \mathbf{x}_{t-1} = \frac{\sqrt{\bar{\alpha}_{t-1}}\beta_t}{1-\bar{\alpha}_t} \hat{\mathbf{x}}_{0|t} + \frac{\sqrt{\alpha_t}(1-\bar{\alpha}_{t-1} )}{1-\bar{\alpha}_t}\mathbf{x}_t + \sigma_t \epsilon,
\end{equation}
with $\epsilon \sim \mathcal{N}(\mathbf{0}, \mathbf{I})$.


Thus far, we have formulated the solution considering the problem without any corruption from noise that can instead occur in the transmission over the channel. Now, we extend the approach to more generic noisy problems of the form $\mathbf{r}=\mathbf{A}\mathbf{x}+\mathbf{n}$ with $\mathbf{n} \sim \mathcal{N}(\mathbf{0}, \sigma_r^2\mathbf{I})$. When applying the operator $\mathbf{A}$ to these problems, a further noisy term $\mathbf{A}^\dagger \mathbf{n}$ is introduced in eq.~\eqref{eq:x0t} as:
\begin{align}
\label{eq:x0t+n}
    \hat{\mathbf{x}}_{0|t} 
    & = \mathbf{A}^\dagger \mathbf{r} + (\mathbf{I} - \mathbf{A}^\dagger \mathbf{A}) \mathbf{x}_{0|t} \cr 
    & = \mathbf{x}_{0|t} - \mathbf{A}^\dagger (\mathbf{A} \mathbf{x}_{0|t}- \mathbf{A}\mathbf{x}) + \mathbf{A}^\dagger\mathbf{n},
\end{align}
leading to distorted or noisy final samples. To make the process aware of the additional noisy term and model the distortion brought from it, we can introduce the parameters $\Sigma_t$ and $\Phi_t$  as:

\begin{align}
    \hat{\mathbf{x}}_{0|t} &= \mathbf{x}_{0|t} - \Sigma_t \mathbf{A}^\dagger (\mathbf{A} \mathbf{x}_{0|t}- \mathbf{r}),\\
    \hat{p}(\mathbf{x}_{t-1}|\mathbf{x}_t, \hat{\mathbf{x}}_{0|t}) &= \mathcal{N}(\bmu_t (\mathbf{x}_{t-1}|\mathbf{x}_t, \hat{\mathbf{x}}_{0|t}), \Phi_t\mathbf{I}).
\end{align}

\noindent The two parameters $\Sigma_t$ and $\Phi_t$ play a crucial role in scaling the range space correction and the noise $\sigma_t\epsilon$ in  $\hat{p}(\mathbf{x}_{t-1}|\mathbf{x}_t, \hat{\mathbf{x}}_{0|t})$, respectively.
However, to ensure the correction properly works, such parameters have to satisfy two constraints:

\begin{enumerate}
    \item To maximize the consistency constraint via the range space correction, $\Sigma_t$ should tend to the identity matrix;
    \item To guarantee that the pretrained model $\mathcal{Z}_\theta$ is able to remove the noise in $\mathbf{x}_{t-1}$, $\Phi_t$ should ensure that the noise variance in $\mathbf{x}_{t-1}$ is equal to the scheduled one $\sigma_t^2$.
\end{enumerate}

By approximating the additive noise in eq.~\eqref{eq:x0t+n} with $\mathbf{A}^\dagger\mathbf{n} \sim \mathcal{N}(\mathbf{0}, \sigma_r^2 \mathbf{I})$, it is possible to simplify the two parameters to be $\Sigma_t = \lambda_t\mathbf{I}$ and $\Phi_t=\gamma_t\mathbf{I}$ \cite{Wang2022ZeroShotIR}. Consequently, we can update the intermediate state $\mathbf{x}_{t-1}$ in eq.~\eqref{eq:interstate} with the constraint satisfaction by setting as follows:
\begin{align}
    \gamma_t &= \sigma_t^2 - \left( \frac{\sqrt{\bar{\alpha}_{t-1}}\beta_t}{1-\bar{\alpha}_t} \lambda_t \sigma_t \right)^2,\\
    \lambda_t &= \begin{cases}
        1, & \mbox{if} \ \sigma_t \ge \frac{\sqrt{\bar{\alpha}_{t-1}}\beta_t}{1-\bar{\alpha}_t}\sigma_{\mathbf{r}} \\
        \sigma_t/ \sigma_{\mathbf{r}}, & \mbox{if}\ \sigma_t < \frac{\sqrt{\bar{\alpha}_{t-1}}\beta_t}{1-\bar{\alpha}_t}\sigma_{\mathbf{r}}
    \end{cases}
\end{align}

\noindent for the first constraint, and:

\begin{equation}
    \left(\frac{\sqrt{\bar{\alpha}_{t-1}}\beta_t}{1-\bar{\alpha}_t} \lambda_t \sigma_{\mathbf{r}} \right)^2 +\gamma_t = \sigma_t^2
\end{equation}

\noindent for the second constraint. The optimal value of the denoising parameter can be directly estimated from the received signal as \cite{Grassucci2023DiffusionMF}:

\begin{equation}
    \sigma_\mathbf{r}^* = (\max(\mathbf{r}) - \min(\mathbf{r})) \cdot \sigma_{\mathbf{r}}. 
\end{equation}

Algorithm \ref{alg:algo1} shows the proposed range-null diffusion sampling steps per user $k$ and the analogies with the multi-user scenarios described in Sec. \ref{sec:background}.


\begin{algorithm}[t]
\caption{\textbf{Null–Space Diffusion Sampling} (per user $k$)}
\label{alg:algo1}
\begin{algorithmic}[1]
\Require Received vector $\mathbf{r}_k$, known channel‑allocation matrix $\mathbf{A}_k$, steps $T$, pretrained neural network denoiser $\mathcal{Z}_\theta$
\State $\mathbf{x}_T \sim \mathcal{N}(0,I)$                                        \Comment{initial noise}
\For{$t = T,\dots,1$}
    \State $\mathbf{x}_{0\mid t} \gets \dfrac{\mathbf{x}_t - \mathcal{Z}_\theta(\mathbf{x}_t,\,t)\sqrt{1-\bar\alpha_t}}{\sqrt{\bar{\alpha}_t}}$
    \Comment{Eq.\,(21)}
    \Statex \hfill\textcolor{gray}{\small// – – range / null split – –}
    \State $\mathit{range} \gets \mathbf{A}_k^{\dagger}\,\mathbf{r}_k$                              \Comment{\emph{\highlight{teal}{received}}}
    \State $\mathit{null}  \gets (\mathbf{I} - \mathbf{A}_k^{\dagger}\mathbf{A}_k)\mathbf{x}_{0\mid t}$      \Comment{\emph{\highlight{orange}{to generate}}}
    \State $\hat{\mathbf{x}}_{0\mid t} \gets \highlight{teal}{$\mathit{range}$} + \Sigma_t\,\highlight{orange}{$\mathit{null}$}$      \Comment{Eq.\,(25)}
    \State Compute $\mu_t,\;\Phi_t$ with Eqs.\,(27)–(29)
    \State $\mathbf{x}_{t-1} \sim \mathcal{N}\!\bigl(\mu_t,\;\Phi_t I\bigr)$
\EndFor
\State \textbf{return} $\hat{\mathbf{x}}_{k}\!\gets\!\hat{\mathbf{x}}_{0\mid 1}$                    \Comment{final regenerated}
\end{algorithmic}
\vspace{-2pt}
\small\textbf{Range–Null Reminder:} $\mathit{range}=\mathbf{A}_k^{\dagger}\mathbf{r}_k$ is received and known at the receiver;  
$\mathit{null}$ is regenerated at every step so that $\mathbf{A}_k\hat{\mathbf{x}}_k=\mathbf{r}_k$ remains satisfied.
\end{algorithm}





\section{Experimental Evaluation}
\label{sec:exp}

To thoroughly assess the applicability and the performance of the proposed framework, we conduct several experiments with different scenarios, datasets, and evaluations.

\subsection{Scenario}


We consider an OFDMA multi-user scenario for experimental evaluation. As mentioned earlier, $N$ subcarriers can be allocated to each user's downlink transmission among a total of $L = K N$ subcarriers. As a result, it is possible to choose $N$ subcarriers with sufficiently high channel gains for each user, as each user has different channel conditions from the others. In other words, a multi-user diversity gain \cite{TseBook05} can be exploited to assign a subset of subcarriers to each user. Furthermore, as the base station knows the channel coefficients, it is possible to perform power control to equalize the effective channel gains. Consequently, it is sufficient to consider the Additive White Gaussian Noise (AWGN) channel model for each user. In addition, for signal modulation per subcarrier, 16 Quadrature Amplitude Modulation (QAM) is employed with SNRs in the range of $\{-10, -5, 0, 5, 10\}$.
The signal is then demodulated at the receivers with 16QAM as well.

\subsection{Comparison Methods}
We compare the proposed method with three well-known approaches. The first approach is based on LDPC in the IEEE 802.11  WiFi standard as a channel code. Here, we do not pay attention to the source compression, therefore LDPC is directly applied to the modulated source image. Going further, since our method is learning-based, we compare it with a Deep learning-based Joint Source-Channel Coding (DeepJSCC) in two different versions. This model comprises a neural encoder at the sender side and a neural decoder at the receiver side. Usually, these two models are trained in an end-to-end fashion. Within the family of DeepJSCC models, we select a promising method that extends these frameworks to an OFDM-based system, which exploits single-tap frequency domain equalization to mitigate the multipath fading channel \cite{yang2021deep}. Two variants of the DeepJSCC-OFDM method have been proposed. The \textit{implicit} one (DeepJSCC-OFDM-I) concatenates the frequency domain pilots and the data symbols together and directly feeds them into the decoder model. Instead, in the \textit{explicit} approach (DeepJSCC-OFDM-E), the authors introduce two residual light-weight neural networks that they call \textit{subnets} to learn residual errors in the channel estimation and equalization.

\subsection{Datasets}
We conduct the evaluation on two different datasets. First, we consider the CelebA-HQ dataset, which comprises $30k$ high-quality images of human faces at resolution $256 \times 256$ and it is widely adopted to test generative models and communication systems. We use this dataset as the benchmark to conduct comparisons with our method and the baselines since some of them were originally trained in the same dataset.
However, we aim at scaling up the possibilities of the proposed approach and we test the performance of our method in large-scale datasets and with dataset-free images. Indeed, we also consider a diffusion model pretrained on ImageNet, which comprises 1 million labeled images collected from flickr, depicting 1000 object categories. Among these samples, a large set of the classes in this dataset are animals, plants, or other nature-related objects. Moreover, many images contain humans. Therefore, it is quite challenging to reconstruct or generate other kinds of objects, such as urban scenes or buildings. Following \cite{Dhariwal2021BeatGANs}, the images are reshaped to $128 \times 128$ for training, while on the testing stage, we reshape the images to $256 \times 256$. While DeepJSCC approaches fail to train on this dataset due to data complexity, the diffusion model successfully completes the training, allowing us to apply the proposed null-space sampling algorithm on it and to evaluate the performance on the ImageNet test set. It is notable to note that our sampling scheme can be plugged into any pretrained diffusion model, enabling the usage of any foundational model trained on any dataset, considerably expanding the possibilities and the application. This is a considerable advantage over DeepJSCC methods that need to be retrained on specific datasets and for specific tasks.
In addition, with the model trained on ImageNet, we perform zero-shot image inpainting on dataset-free images, which are images collected directly from our camera and that are not contained in any public datasets. To test the generalization ability of the proposed model, we select two extremely challenging images, full of details and containing under-represented classes, that may be urban scenes or buildings. The two selected images are reported in the first column of Fig.~\ref{fig:fig1}.

\begin{figure*}
    \centering
    \includegraphics[width=\textwidth]{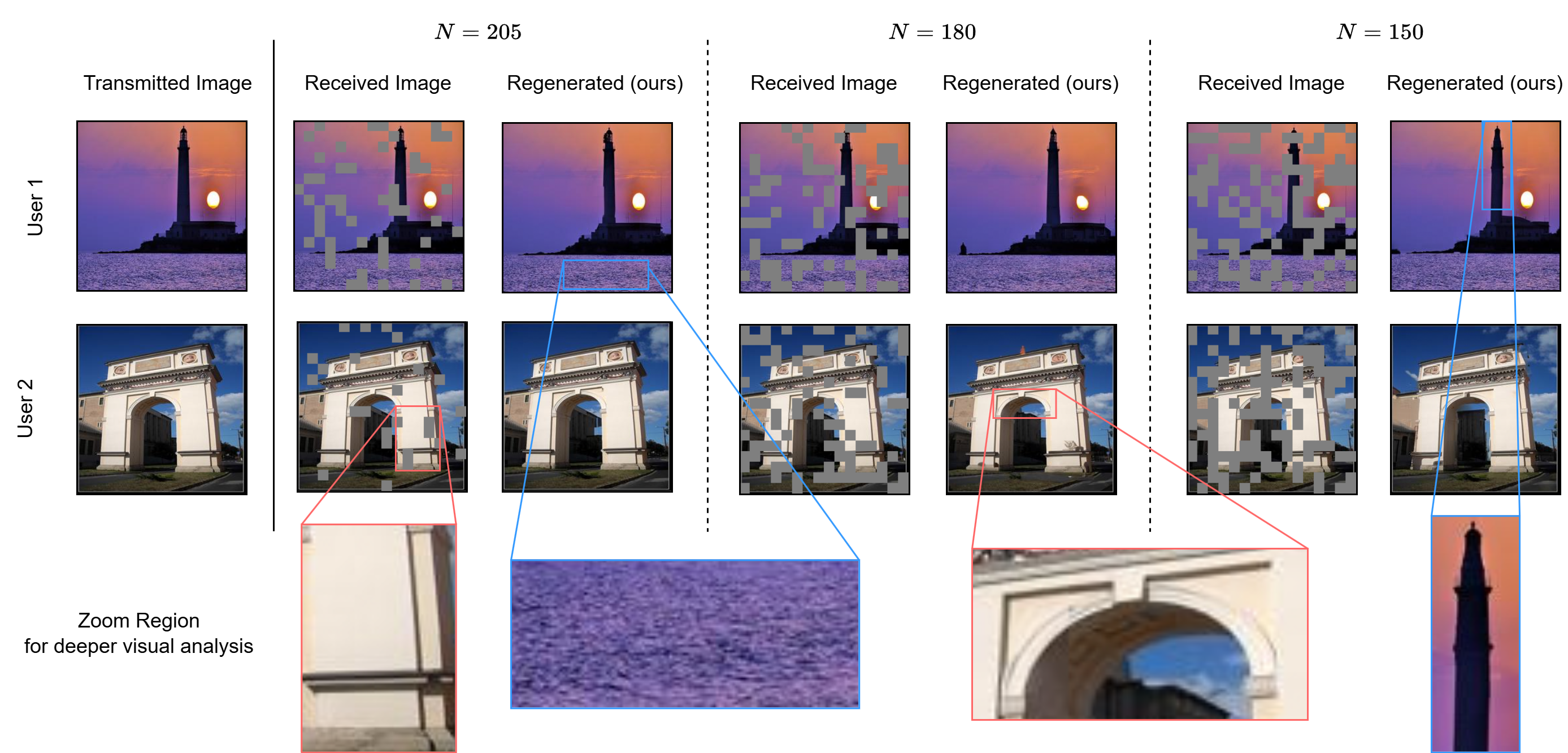}
    \caption{Random samples from the ImageNet test set regenerated by our method. We consider two users and three different channel assignment scenarios, where the total number of subcarriers $M$ is set to $256$, and we experiment with $N < M$ and $N = \{ 205, 180, 150 \}$, equivalent to an $N/M$ ratio of $\{0.8, 0.7, 0.6 \}$, respectively. We also zoom four heavily affected regions for a better visual evaluation. Here, it is clear how our method excellently regenerated missing portions of data even though it receives extremely degraded information.}
    \label{fig:imagenet_results}
\end{figure*}

\begin{figure}
    \centering
    \includegraphics[width=\linewidth]{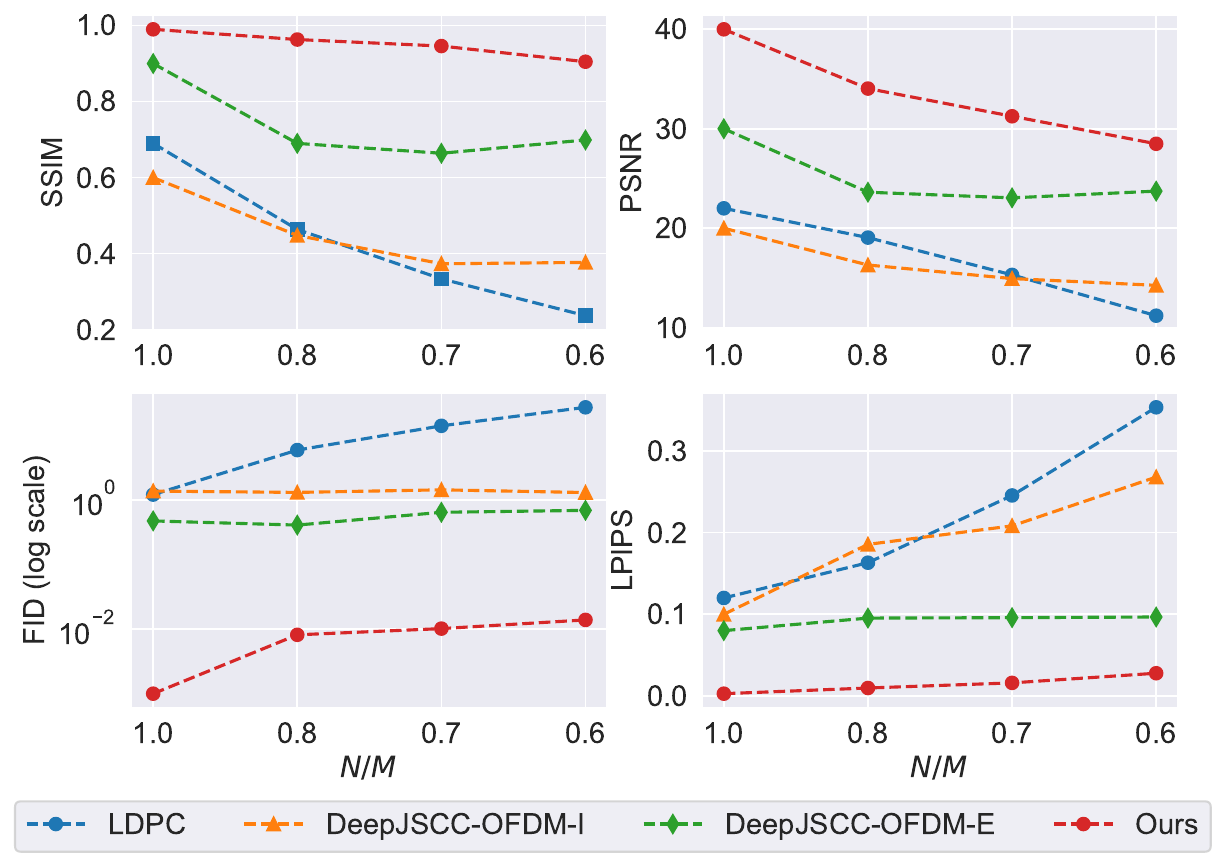}
    \caption{Comparison for different $N/M$ subcarriers ratio, where $N < M$ and $M = 256$, for $K=2$ users during transmission, evaluated with different metrics, namely SSIM$\uparrow$, PSNR$\uparrow$, FID$\downarrow$, LPIPS$\downarrow$ on the CelebA-HQ dataset. The proposed method (in red line) far exceeds any other method according to all the four metrics in each scenario.}
    \label{fig:metrics_comparison}
\end{figure}

\begin{figure*}
    \centering
    \includegraphics[width=\textwidth]{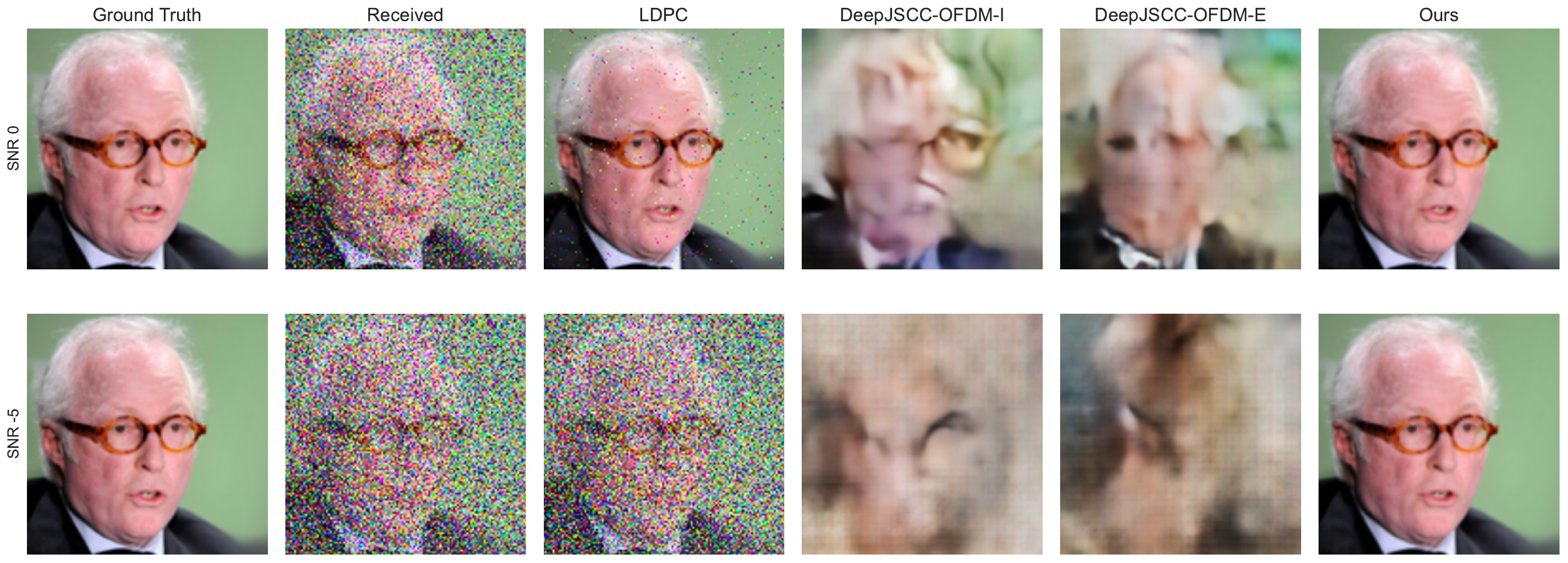}
    \caption{Visual comparison of different methods under extremely low channel noise SNR values ($0$, $-5$).}
    \label{fig:snr_visual_res}
\end{figure*}

\begin{figure}
    \centering
    \includegraphics[width=\linewidth]{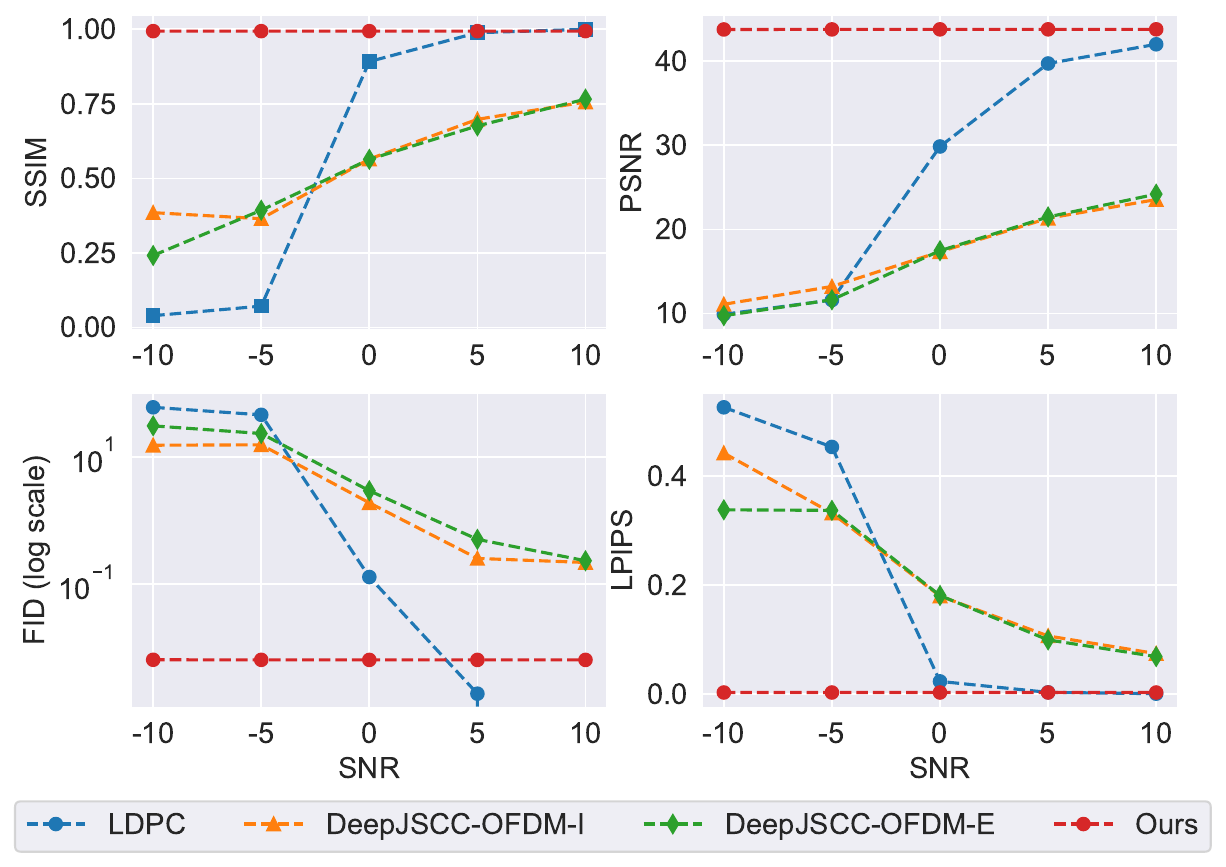}
    \caption{Comparison of the methods against different channel SNR, evaluated with different metrics, namely SSIM$\uparrow$, PSNR$\uparrow$, FID$\downarrow$, LPIPS$\downarrow$ on the CelebA-HQ dataset. These results prove that the proposed method is extremely robust not only to lower numbers of subcarriers but also to different and aggressive channel noise corruption.}
    \label{fig:snr_comparison}
\end{figure}

\subsection{Metrics}
We evaluate the performance of the proposed method under four distinct metrics. As a first evaluation, we involve the common Structural Similarity Index (SSIM), the higher the better, and the Peak Signal-to-Noise Ratio (PSNR), the higher the better as well. However, these metrics assess the performance pixel-by-pixel or bit-per-bit without effectively evaluating whether the semantics of the content have been preserved at the receiver. Therefore, while both SSIM and PSNR are suitable for evaluating conventional communication systems, they are being progressively discarded in the evaluation of semantic communication systems \cite{Mahgoub2026Access, Grassucci2023GenerativeSC}. A more appropriate metric to evaluate the semantic transmission of images is the Learned Perceptual Similarity (LPIPS) \cite{zhang2018perceptual}, which has been demonstrated to be much more aligned with human perception with respect to SSIM and PSNR \cite{mao2023extreme}. The LPIPS metric comprises complex and fine-tuned steps with the aim of evaluating the distance of features in different neural networks. At a low level, let us consider a simple convolutional layer and two generic images in $\mathbb{R}^{C\times H \times W}$, with $C$ being the channel dimension, and $H$ and $W$ the height and width. The LPIPS is computed by evaluating the cosine distance between the two images (in the channel dimension, e.g., the RGB channels for colored images) and averaging across spatial dimensions $H$ and $W$, as well as on the layers of the network. The lower the LPIPS value, the more perceptually similar are the two samples. In our work, as the baseline network to extract features, we consider the conventional VGG model. Finally, we consider a further metric that is currently the most widely adopted for evaluating generative models performance in the image domain. The Fr\'echet Inception Distance (FID) relies on the Fréchet distance between the features extracted from the InceptionV3 network of the original and of the regenerated images. While SSIM measures image degradation through structural information, FID estimates how the distributions of original and regenerated images are far from each other. Therefore, lower FID values correspond to more plausibly regenerated samples. Formally, considering the two sets of features coming from the real ($r$) and the generated ($g$) sets normally distributed, the FID assumes the following form:

\begin{equation}
    \text{FID} = \| \mu_r - \mu_g \|^2 + Tr \left(\Sigma_r + \Sigma_g - 2 \left(\Sigma_r \Sigma_g \right)^{1/2} \right),
\end{equation}

\noindent whereby, $\mu$ represents the mean and $\Sigma$ the covariance matrix of the two Gaussian distributions.

\subsection{Neural Model}
We consider two backbone models for the proposed null-space diffusion model framework. Importantly, the proposed sampling algorithm can be adopted with any pretrained diffusion model, as it is training-independent. To this end, we do not retrain any model and we select two pretrained models that can fit our scenario. For the CelebA-HQ dataset, we employ the pretrained model of \cite{meng2022sdedit}, while for ImageNet we involve the pretrained model from \cite{Dhariwal2021BeatGANs}. For both architectures, the backbone model is a residual U-Net, with global attention layers at three different resolutions, that are $32 \times 32, 16 \times 16, \text{and } 8 \times 8$ with $64$ channels per each of the 4 heads. The two residual blocks for upsampling and downsampling the activations are inspired from \cite{Brock2018LargeSG}, as suggested in \cite{song2020score}. Similarly, the residual connections are scaled by $1 / \sqrt{2}$. Moreover, adaptive group normalization is used for injecting forward and reverse process timesteps into residual blocks. For training, the number of diffusion steps $T$ is set equal to $1000$, with a linear noise schedule.

\subsection{Results}
To provide an exhaustive evaluation, we report both objective and subjective evaluations of the experiments. We conduct experiments to validate the ability of the proposed models to regenerate the data lost during transmission due to $N < M$ subcarriers, then the capabilities of restoring samples from channel noise, and finally we consider both the scenario together.

\textbf{Fill the gaps.} Firstly, we evaluate the performance of the proposed method in regenerating the missing portions of data lost due to the constraint $N < M$ in the number of subcarriers as stated in Sec.~\ref{sec:background}. Figure~\ref{fig:metrics_comparison} displays the value of SSIM, PSNR, FID, and LPIPS under the different values of $N$, say $256, 205, 180, \text{and } 150$, corresponding to a fraction $N/M$ equal to $\{1, 0.8, 0.7, 0.6\}$ on the CelebA-HQ dataset. The LDPC method, in the blue line, is unable to face the missing data case, and its performance drastically sinks as the $N/M$ ratio decreases. On the contrary, the proposed method, in the red line, preserves high performance across each situation, clearly outperforming all other comparisons. Indeed, our method barely shows any loss in LPIPS and FID scores from $0.2$ loss to $0.6$ subcarriers lost, due to its potential to inpaint missing portions of images. This result proves the capability of our method as a crucial component of future multi-user OFDMA systems, in which the amount of information to transmit may be strictly related to the channel state and to the ability of the null-space diffusion model to regenerate the missing portions of data. Moving to a subjective evaluation, Figure~\ref{fig:imagenet_results} shows random samples from the ImageNet test set with received and infilled images for a scenario with two users and $ N/M \text{ in }\{0.8, 0.7, 0.6\}$, corresponding again to $N=\{ 205, 180, 150\}$. Despite several chunks of lost information, the proposed method excellently fills the gaps to be semantically consistent with the whole picture and with the neighbors of the gaps. Even when the gaps almost completely cover some objects in the picture (i.e., the lighthouse in the last picture, or the arch decoration in the middle one), the proposed method inpaints the missing information in order to be extremely consistent with the neighbor content. Moving forward, additional results that shows the ability of our method are shown in Fig.~\ref{fig:fig1}. Here, it is notable that the tested images are dataset-free and are not contained in any public online dataset, as they are pictures directly taken by the authors of this work. However, even in this very challenging scenario, the proposed method is able to restore the image and fill the gaps with semantic-consistent content.

\begin{figure}
    \centering
    \includegraphics[width=\linewidth]{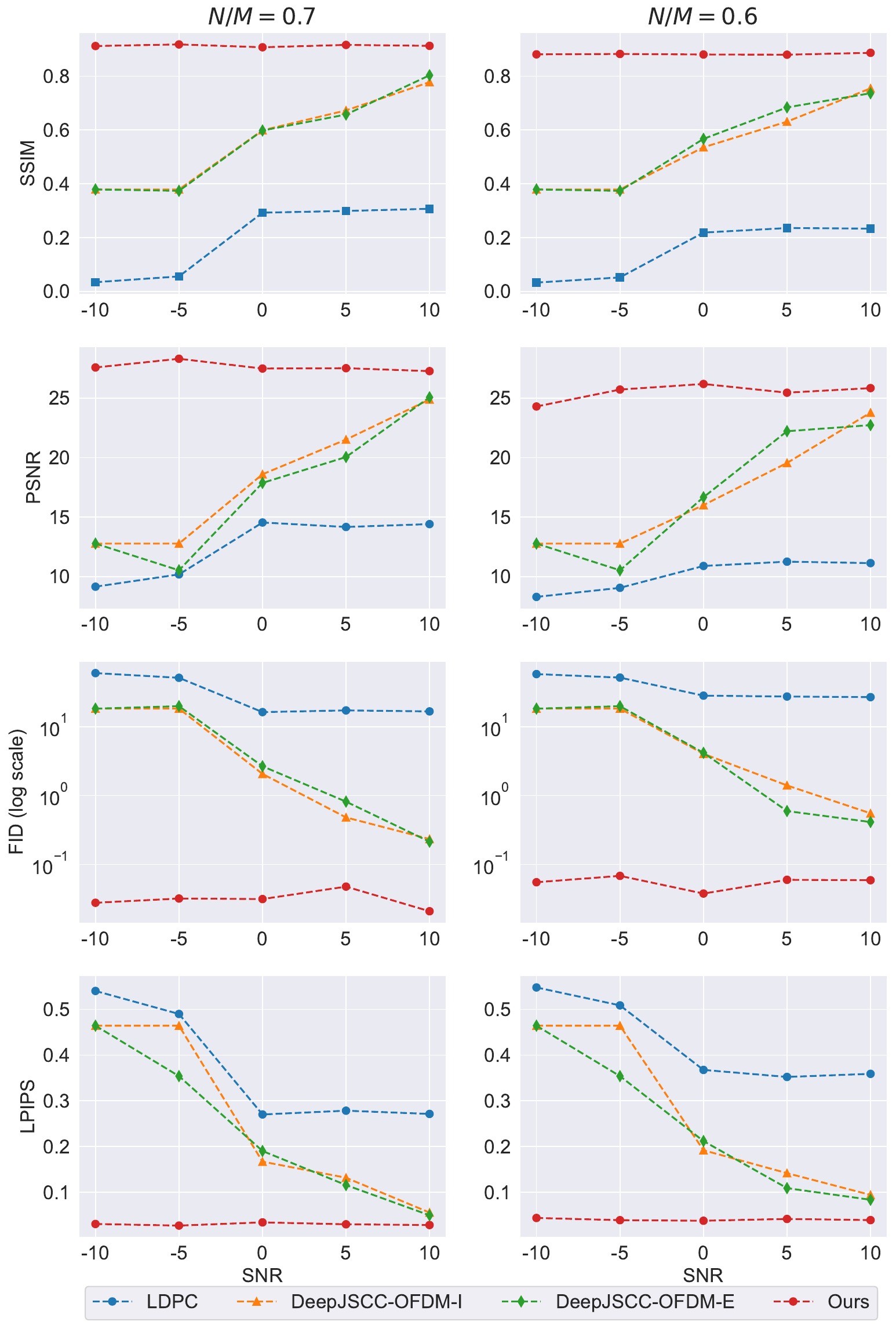}
    \caption{Results for the experiments with joint $N < M$ subcarriers and different channel noise SNR values. We consider two scenarios for the $N/M$ ratio equal to $0.7$ and $0.6$ in the first and second columns, respectively. For both the scenarios we consider an SNR equally spaced from $-10$ to $10$. Our method, in red, largely outperforms all other comparisons in each of the scenarios considered.}
    \label{fig:snr&n_ratio_results}
\end{figure}

\begin{figure}[t]
    \centering
    \includegraphics[width=\linewidth]{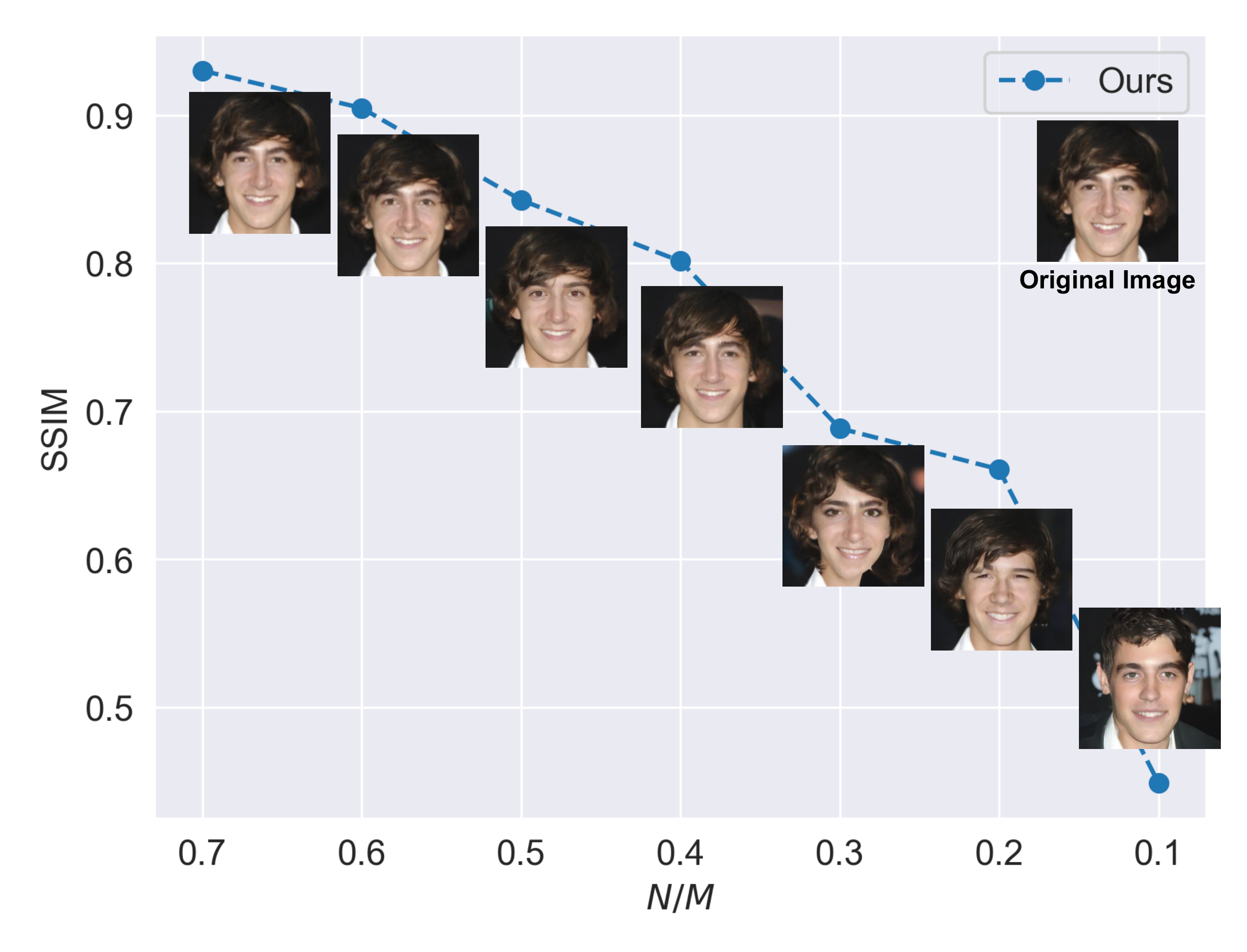}
    \caption{Performance of the proposed method in extreme scenarios with very low $N/M$ ratios, from $0.7$ to $0.1$.}
    \label{fig:stress_test}
\end{figure}

\textbf{Denoise.} Together with the ability to inpaint image gaps, we also test the robustness of the proposed method under different noise channel SNRs. We consider extremely strong conditions and evaluate the performance with SNR values within $\{-10, -5, 0, 5, 10\}$. As Figure~\ref{fig:snr_comparison} highlights, the proposed method far exceeds any other comparison whose performance drastically drops the lower the SNR becomes. For each metric we consider, our model preserves the performance by incomparably denoising and regenerating corrupted samples at the receiver. Indeed, thanks to the iterative denoising of the sampling procedure as explained in Subsection~\ref{subsec:method}, the null-space diffusion model can effectively denoise the received images even in the case of extremely low SNR values. This result is also strengthened by the visual comparison we report in Fig.~\ref{fig:snr_visual_res}, in which a random test sample from the CelebA-HQ dataset is transmitted with the challenging SNR equal to $0$ and $-5$. The received image is highly corrupted by the AWGN channel noise, and most of the existing approaches fail to recover the ground truth (original) transmitted sample. On the contrary, our method excellently restores the corrupted sample producing a visually-pleasant regenerated image.

\textbf{Altogether: denoise \& fill the gaps.} Lastly, we test the capabilities of the proposed method to both denoise and fill the missing data portions altogether. We select two $N/M$ ratios, which are $0.7$ and $0.6$, and for each one with simulated experiments with extremely low SNR values, equally spaced from $-10$ to $10$. Figure~\ref{fig:snr&n_ratio_results} shows the results for each scenario and the three comparison methods against our approach. From the curves in Fig.~\ref{fig:snr&n_ratio_results}, it is once more clear that the proposed method is robust against both aggressive channel noise corruptions and low numbers of subcarriers with $N$ strictly lower than $M$. Indeed, according to the four objective metrics (SSIM, PSNR, FID, and LPIPS), our method keeps high performance across the different scenarios, while other methods progressively fail.

In conclusion, the experimental evaluation undoubtedly proves the efficacy of the proposed null-space diffusion model as a novel generative-based framework for multi-user semantic communication. Thanks to the null-space sampling procedure, the method can effectively solve communication inverse problems and fill the gaps due to the multi-user scenario in received information, as well as denoise the content from extremely aggressive channel noise.

\subsection{How low can the $N/M$ ratio be for our model?}

We test the proposed method in extreme conditions, including all the possible $N/M$ ratios up to $0.1$ to understand the furthest capabilities of our approach. Figure~\ref{fig:stress_test} shows the SSIM curve and the respective regenerated images for $N/M$ ratios equally spaced from $0.7$ to $0.1$. From the curve and the visual analysis, it is clear how our method still regenerates plausible samples up to $N/M = 0.4$, which means only $102$ subcarriers over $256$, with an SSIM equal to $0.8$. Moreover, the performance is acceptable also up to an $N/M$ ratio of $0.2$, in which the system is using only $51$ subcarriers among the total $256$, obtaining an SSIM of $0.66$ and recognizable regenerated image. The performance consistently drops only at the ultimate ratio of $0.1$ ($26$ subcarriers).
In conclusion, this experiment shows how the proposed method is exceptionally robust to extremely low number of employed subcarriers, proving once more the capability of our model to be involved in multi-user scenarios.


\section{Conclusions and Future Works}
\label{sec:con}

In this paper, we introduced and formulated a novel generative model-based semantic communication framework for multi-user scenarios. The proposed null-space sampling technique regenerates each user's missing signals, which have been deliberately not sent to save radio resources for other users. It was also shown that the proposed technique can concurrently correct noisy channel perturbations.  Across a wide range of SNRs, including extremely low SNR regimes, our proposed diffusion model-based framework has consistently achieved higher perceptual similarities than classical LDPC and autoencoder-based semantic communication frameworks such as DeepJSCC. 

Extending this preliminary study, where full channel knowledge at the transmitter is assumed, to incorporate the impact of channel estimation could be an interesting topic for future research. 
Another interesting direction could be applying diffusion sampling acceleration techniques like quantization to reduce computing latency and complexity at the receiver. Nevertheless, note that the proposed sampling algorithm can be plugged into any other pretrained diffusion model, so this extension is straightforward.




\bibliographystyle{ieeetr}
\bibliography{MUSCBiblio}


\end{document}